# Nonlinear Nonlocal Metasurface for Harmonic Generation and Manipulation

Hooman Barati Sedeh[1], Yuruo Zheng[1], Jiaren Tan[1], Luca Carletti[2], Danilo Gomes Pires[1], Anna R. Finkelstein[1], Maria Antonietta Vicenti[2], Ivan Kravchenko[3], Michael Scalora[4], and Natalia M. Litchinitser[*,1]

[1] Department of Electrical and Computer Engineering, Duke University, 27708 Durham, NC, USA.

[2] Department of Information Engineering, University of Brescia, Via Branze 38, 25123, Brescia, Italy.

[3] Center for Nanophase Materials Sciences, Oak Ridge National Laboratory, Oak Ridge, TN 37831, USA.

[4] Aviation and Missile Center, US Army CCDC, Redstone Arsenal, AL 35898-5000, USA.

[*]Corresponding Author: natalia.litchinitser@duke.edu

**Abstract**

The discovery of the second-harmonic generation in 1961 marked the birth of nonlinear optics and enabled a plethora of applications, ranging from harmonic generation to quantum light sources. Despite these advances, achieving phase matching in bulk nonlinear crystals remains a key bottleneck for achieving efficient photon–photon interaction. While thinning nonlinear materials relaxes phase-matching constraints, it severely limits the efficiency of nonlinear optical processes, due to the drastically shortened interaction length. Although photonic metasurfaces—planar arrays of subwavelength meta-atoms—offer a promising alternative due to their ability to support resonant modes, enabling relatively strong local field enhancement, existing designs face a trade-off between high nonlinear conversion efficiency, enabled by nonlocal metasurfaces, and advanced wavefront control facilitated by local ones. These two capabilities currently remain fundamentally decoupled due to the fundamental differences between local and nonlocal regimes. To efficiently utilize the advantages of both regimes, we strategically design a nonlinear nonlocal metasurface that supports quasi-trapped modes (QTM), enabling both efficient harmonic generation and phase manipulation at the meta-atom level. By engineering an all-dielectric metasurface with topologically asymmetric meta-atoms, we achieve strong field confinement and demonstrate third-harmonic generation with an enhancement of over three orders of magnitude compared to unstructured thin films. Utilizing the structure's $C_1$ symmetry and Pancharatnam–Berry (PB) phase manipulation via meta-atom rotation, we demonstrate helicity-dependent wavefront control at both fundamental and third-harmonic wavelengths. We further show that a slight perturbation to the meta-atom geometry results in geometric phase accumulation exclusively at the resonant wavelength—a behavior not observed in conventional linear and nonlinear PB-based metasurfaces. This selectivity originates from the unique QTM field profile, which preserves global symmetry off-resonance while enabling local geometric phase encoding at resonance for both fundamental and harmonic signals, thereby modifying the conventional selection rules previously assumed under circularly polarized illumination. These results not only expand the functional scope of silicon photonics for advanced optical communication and quantum information systems but also uncover new physical mechanisms in nonlinear geometric phase control at the nanoscale.

**Introduction**

The advent of photonic metasurfaces (MS), engineered interfaces composed of subwavelength meta-atoms arranged in spatially engineered patterns, has provided an exceptional platform for controlling the amplitude, phase, and polarization of optical fields with subwavelength precision in both spatial and temporal domains [1-4]. Two major classes of these nanoscale platforms include local and nonlocal metasurfaces [5-7]. In particular, a fundamental principle for designing local MSs is the assumption that the scattering response at a specific point is unaffected by the light-matter interactions at other points [5]. This assumption facilitates the straightforward design of MS structures, where each meta-atom is chosen from a library of meta-atom geometries. Such MSs usually possess a narrow optical response in the real space, corresponding to a broad response in reciprocal space. As a result, local MSs usually possess broadband and angle-insensitive response [5-7]. In contrast, in nonlocal MSs, an optical response at any given point is influenced by that of the meta-atoms at distant locations along the surface of the structure. As a result, the broader spatial extent of scattering in real space leads to a narrow response in reciprocal space, yielding spectral and angular selectivity [5]. Although extensive research has been devoted to studies of light-matter interactions in both local and nonlocal MSs in the linear optical regime [8-18], recent studies have increasingly focused on nonlinear optical phenomena within these two-dimensional subwavelength structures [19-26]. This rapidly growing research area has enabled a range of advanced functionalities, including nonlinear harmonic generation and wavefront manipulation [27-32], the creation of entangled photon pairs [33-38], and the generation of nonlinear structured light beams [39-43]. However, while nonlocal MSs offer the advantage of higher nonlinear conversion efficiencies due to their high Q-factor modes that spatially extend across multiple meta-atoms, these resonant modes fundamentally constrain the degree of control over the spatial shape of the generated optical wavefront. As a result, the simultaneous generation of nonlinear harmonics and their spatial wavefront manipulation remains a fundamental challenge for nonlocal MSs.

To overcome this limitation, in this work, we combine the advantages of both local and nonlocal MSs and demonstrate a novel design of an ultrathin silicon nonlocal MS supporting quasi-trapped resonant modes (QTM), as illustrated in **Fig. 1a**. Owing to the topology of the constituent meta-atoms, shown in the inset of Fig. 1a (right), with $D$ being meta-atom diameter, $H$ is its height, $r_h$ is the radius of the hole, and $\Delta$ being the displacement of the center of the hole with respect to the center of the meta-atom, the resulting electric field is strongly confined both at the surface and within the bulk of the structure. In particular, the introduction of the nanoscale hole not only enhances the electric fields within the bulk of structure but also leads them to rotate in the transverse plane of the meta-atoms as shown in Fig. 1a (left) resulting in the THG with over three orders of magnitude enhancement compared to unstructured thin films (Fig. 1b). The introduction of the off-center hole breaks the $C_\infty$ rotational symmetry of the pillars to $C_1$, such that under circular polarization (CP) excitation, spatially-rotating each meta-atom with an angle $\alpha$ imparts a continuously varying geometric phase shifts of $2\alpha$ and $4\alpha$ on the co- and cross-polarized third-harmonic components, respectively (Fig. 1c) [19]. As a result, the presented MS design offers full $2\pi$ phase coverage and helicity-dependent control over the third-harmonic wavefront, while simultaneously enabling high conversion efficiency due to the resonant nature of the structure.

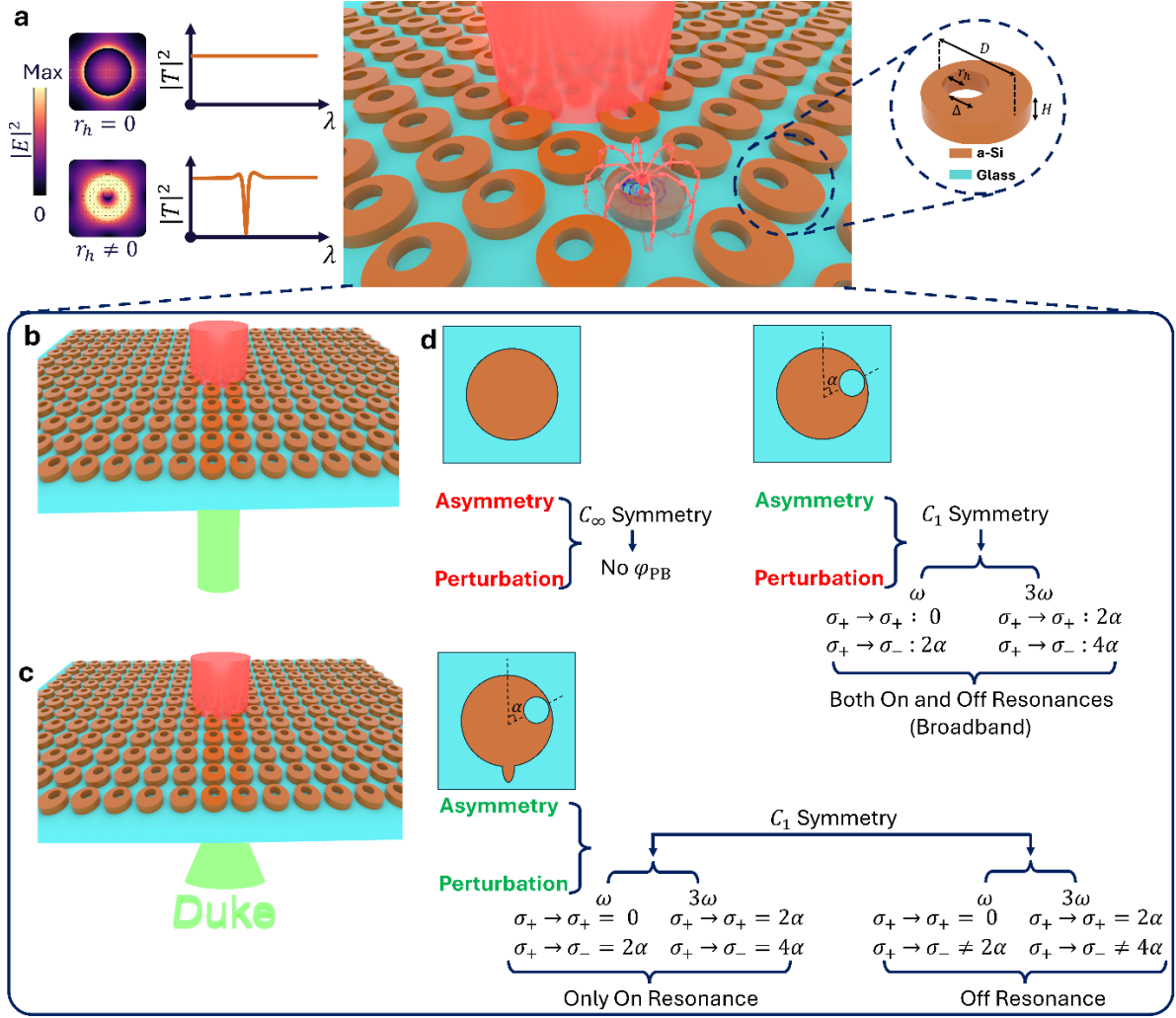

**Fig. 1 | Schematic illustration of the nonlocal nonlinear harmonic generation and manipulation. a**, Design of the ultrathin silicon nonlocal metasurface supporting quasi-trapped resonant modes. Strong electric field confinement occurs both at the surface and within the volume of the structure due to the engineered topology of the constituent meta-atoms (left). The inset shows a single meta-atom with design parameters of $D$ (diameter), $H$ (height), $r_h$ (hole radius), and $\Delta$ (offset of the hole from the center). **b**, Conceptual depiction of THG generation from the metasurface, which experimentally yields over three orders of magnitude enhancement compared to an unstructured thin film under low incident pump power. **c**, Visualization of helicity-dependent wavefront manipulation enabled by nonlinear geometric phase control. The introduction of an off-center hole breaks the $C_\infty$ symmetry to $C_1$, allowing each meta-atom rotated by an angle $\alpha$ to impart nonlinear phase shifts of $2\alpha$ and $4\alpha$ on the co- and cross-polarized third-harmonic components, respectively, enabling full $2\pi$ nonlinear phase coverage and generation of complex structured harmonic beams under both right- and left-hand circular polarizations. **d**, Schematic of the modified meta-atom incorporating a nanoscale elliptical hole. The slight perturbation confines the imparted geometric phase to the resonant wavelength in both linear and nonlinear regimes, leaving off-resonant responses unaffected by the rotation of the meta-atom. The behavior arises from the unique field distribution of the quasi-trapped mode, which is concentrated around the hole at resonance, enabling independent phase control at resonant and nonresonant wavelengths—demonstrated here for the first time in the nonlinear regime.

Notably, the geometric phase is generally independent of wavelength and, as such, does not distinguish between resonant and nonresonant regimes, thereby limiting its applicability in narrowband devices. In this work, we overcome this limitation by slightly modifying the meta-atom geometry through the addition of a nanoscale elliptical perturbation, as shown in Fig. 1d. This confines the geometric phase to the resonant wavelength in both linear and nonlinear regimes. As a result, the off-resonant response becomes insensitive to meta-atom rotation, and entirely new nonlinear PB selection rules appear in the TH spectrum. We discuss the origin of this unique behavior, arising from the distinct field distribution of the QTM modes both on and off resonance. To the best of our knowledge, this work is the first to demonstrate that, in a judiciously designed nonlocal metasurface, at resonant wavelengths, geometric phases accumulate in both linear and nonlinear regimes. These phases are governed by the meta-atom geometries and the resulting internal field distributions. In contrast, no geometric phase is accumulated at nonresonant wavelengths, allowing for wavelength-dependent control of the phase under both resonant and nonresonant conditions for linear and nonlinear excitations. Based on the proposed design, we demonstrate the generation and manipulation of various TH wavefronts for both right- and left-handed circularly polarized excitation, as schematically illustrated in Fig. 1c. The unique combination of geometric phase accumulation limited to resonant wavelengths and strong field enhancement establishes a versatile platform for nonlinear, nonlocal wavefront engineering, paving the way for generating complex structured beams at harmonic frequencies.

## Results

### Linear response of a nonlocal metasurface

We design several metasurfaces composed of a-Si pillars on a silicon dioxide substrate, whose scattering properties are controlled by introducing an off-center hole inside the meta-atoms with different radii, as shown in Fig. 1a. Meta-atoms are carefully optimized to exhibit resonant behavior under x-linearly polarized (LP) illumination (see **Methods** and Supplementary Note 1 for more details on the simulation results) and to gain simultaneous significant nonlinear conversion efficiency and desired wavefront of the generated TH. The optimization is based on full-vectorial electromagnetic simulations using the finite-element method implemented in COMSOL Multiphysics. **Figure 2a** illustrates the linear transmission spectra as functions of wavelength, air-hole radii ($r_h$), and off-center distance (Δ) with the metasurface period fixed to $p = 1000$ nm. When the $C_\infty$ symmetry of the meta-atom is not broken (Δ = 0), the degenerate trapped modes do not couple to the field of incoming radiation as shown in Fig. 1a (left). As a result, no resonant modes are excited. To achieve the necessary coupling between incoming radiation and the resonant mode, structural asymmetry is introduced by introducing an off-center hole, which leads to the excitation of quasi-trapped modes with finite and controllable Q-factors [44-47]. Upon breaking the topological symmetries of the meta-atoms, trapped modes leak into the free space and their radiative Q-factors become proportional to both air-hole size and off-center distance, as manifested by the variation of linewidths of the resonant modes within the transmittance spectra (Fig. 2a). To experimentally test these predictions, we fabricated two sets of samples with a fixed periodicity of $P = 1000$ nm and diameters of the meta-atoms being 610 nm, but with varying air-hole radii and off-center displacements of the air-hole, using electron beam lithography (see **Methods**). For both sets, the air-hole radius varied from $r_h = 50$ nm to $r_h = 130$ nm in steps of 20 nm, while the off-center distance was fixed at $\Delta_1 = 100$ nm (first sample) and $\Delta_2 = 50$ nm (second sample), respectively. The scanning electron microscopy (SEM) images of the samples, along with their corresponding linear transmittance spectra measured using a bright-field transmission microscopy setup, are shown in Fig. 2b (see **Methods**).

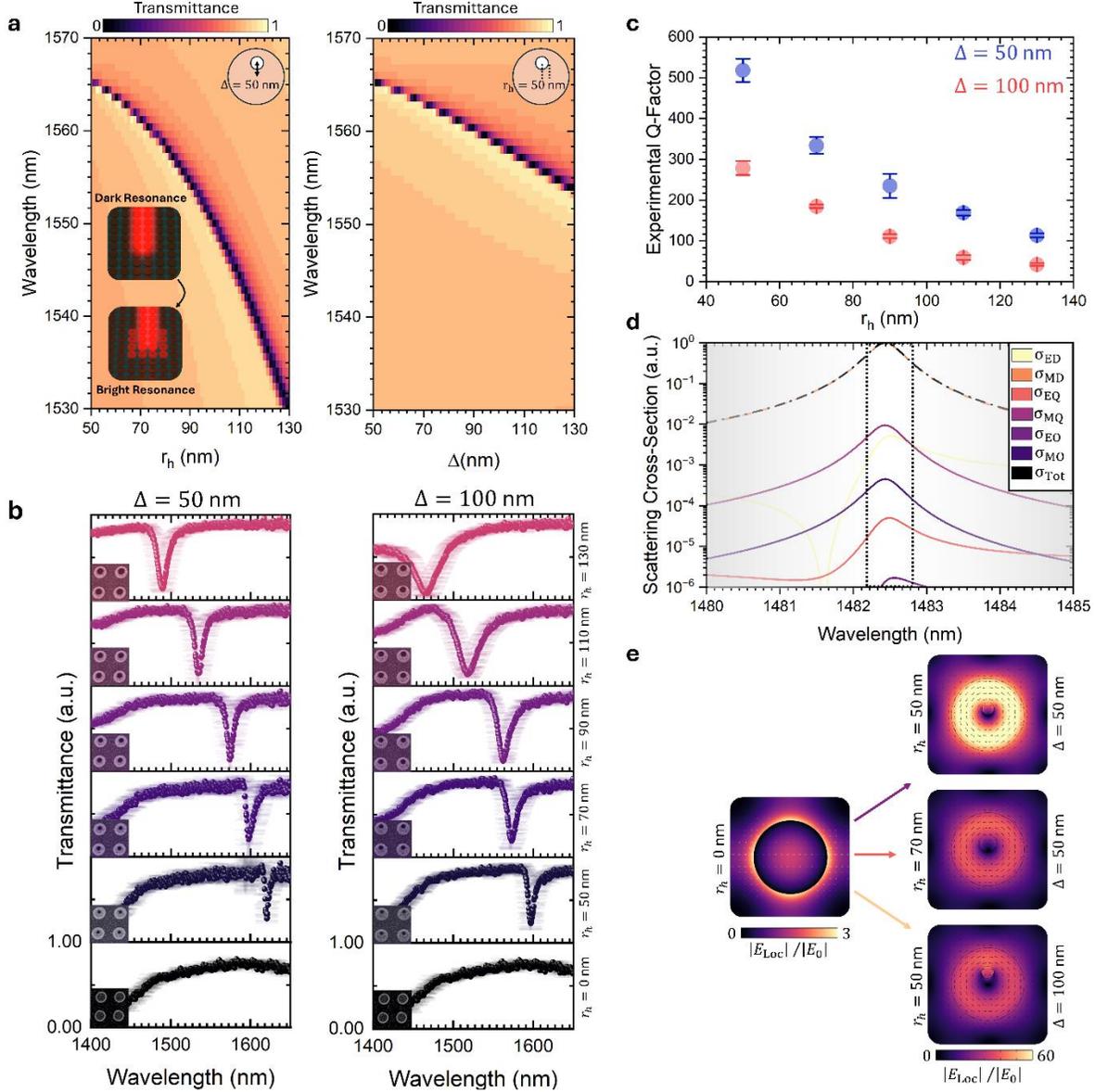

**Fig. 2 | Design and linear characterization of nonlocal metasurface. a**, Calculated numerical simulation of transmittance spectra of the designed metasurface as a function of air-hole radius (when $\Delta = 50$ nm) and off-center distance (when $r_h = 50$ nm). As the in-plane asymmetry of the meta-atom increases, the resonance linewidth increases while blue shifting toward shorter wavelengths. **b**, The measured transmittance spectra of the fabricated samples for meta-atoms with $\Delta = 50$ nm (left column) and $\Delta = 100$ nm (right column) as air-hole radius changes from $r_h = 0$ nm to $r_h = 130$ nm. Once the symmetry is broken by introducing the nanoscale holes, the dark mode begins to leak into the continuum. **c**, Experimental Q-factor of the samples measured by fitting the transmittance spectra using TCMT. The measured Q-factor can reach as high as 500, for $(r_h, \Delta) = (50, 50)$. **d**, The calculated scattering spectra of the designed meta-atoms (neglecting substrate effect), demonstrating dominant magnetic response contributing 98% of the total scattering. The out-of-plane component of the magnetic dipole is the sole contributor in its respective channel, while the in-plane components of the magnetic quadrupole and electric dipole moments contribute to their scattering. **e**, Electric field distribution within the designed meta-atoms when the asymmetry parameter changes. The maximal field enhancement decreases as the asymmetrical factor increases, abiding well with the broadening of the resonant linewidth within the transmittance spectra.

Breaking the $C_\infty$ symmetry of the meta-atom leads to resonant dips within the measured transmittance spectra that blueshift toward shorter wavelengths as the asymmetric factors ($r_h$ and $\Delta$) increase. Such a spectral transition is accompanied by the broadening of the resonant linewidth as the asymmetry parameters change, indicating the reduction in the achievable radiative Q-factor. To experimentally quantify the Q-factor of the fabricated MSs, the transmittance spectra were fitted using temporal coupled mode theory (TCMT), as shown in Fig. 2c (see Supplementary Note 2 for further details and a comparison with existing literature). Next, to understand the underlying physics of the emerged resonant mode and demonstrate the nonlocality of the proposed metasurface, the spherical multipole expansion theory [48-50] was used to calculate the contribution of each optical multipole to the far-field scattering around the trapped mode wavelength for both isolated and periodic array of meta-atoms having $r_h = 70$ nm and $\Delta = 50$ nm, as shown in Fig. 2d (see **Methods**). In particular, at the resonant wavelength of ~1482 nm, a significant peak is seen for the case corresponding to the periodic array. This peak is dominated by a magnetic dipole, accounting for 98% of the total power, with a minor additional contribution of 2% resulting from the combination of magnetic quadrupole and electric dipole effects. In contrast, for an isolated meta-atom (without the nonlocal contribution provided by the coupling of adjacent meta-atoms), only local Mie-type resonant modes are excited, which results in a scattering cross-section that is 1843 times smaller than that of the array configuration, dominated by the nonlocal contributions (see Supplementary Note 3 for more details). Note that the observed spectral shift of the QTM resonant mode from 1560 nm to 1482 nm is due to the absence of the glass substrate in the former case, as discussed in detail in Supplementary Note 3. The out-of-plane component of the magnetic dipole is the sole contributor to scattering in its respective channel, whereas the in-plane components of the magnetic quadrupole and electric dipole moments dominate their corresponding contributions. This interplay among different multipolar moments generates distinct electric and magnetic field distributions within both the surface and bulk regions of the meta-atoms, as illustrated in Fig. 2e for varying air-hole radii, with white arrows indicating the electric field vectors. In the symmetric configuration, although some mode leakage occurs at the meta-atom boundaries, the electric field strength remains substantially weaker throughout the structure compared to the case with an off-center air hole, where the asymmetry significantly enhances the field intensity.

While the geometric setup we propose is similar to the one studied by Matsudo et al. [27] for generating and controlling nonlinear harmonic signals, our method introduces a clear and significant difference. In the earlier work, two resonant modes, one broad local resonance (magnetic dipole) and another nonlocal QBIC, coexisted closely in the spectral domain. Although theoretical and numerical analyses in [27] predicted that the nonlocal mode would dominate the nonlinear response, this prediction has not been observed experimentally. Instead, the measured harmonic signals resulted primarily from local resonance, similar to the results reported by several other groups. In contrast, we judiciously engineered our structure to support a single, spectrally isolated resonant mode of purely nonlocal origin, thereby explicitly eliminating the contributions of any local resonances in its spectral vicinity. This carefully designed approach enabled us to directly observe and experimentally confirm the nonlinear response associated solely with the nonlocal resonance. Consequently, our measured nonlinear signals and the corresponding wavefront manipulations can be uniquely attributed to the nonlocal resonance.

**Non-local nonlinear interactions**

To achieve a strong nonlinear response, any nanophotonic-based platform must not only generate local field enhancement by exciting resonant modes at all wavelengths involved in the nonlinear process, but also ensure that these modes exhibit significant spatial overlap to maximize the efficiency of their interactions within the nonlinear material volume [51-54]. While most existing studies have focused on satisfying the first condition in the context of nonlinear metasurfaces, less attention has been given to the latter one, as achieving both conditions simultaneously is a challenging task. In our design, the rotating nature of the fundamental frequency (FF) fields, as shown in Fig. 2e, induces transverse components of the nonlinear polarization $\boldsymbol{P}_{nl}(3\omega)$, resulting in a third-harmonic (TH) field that exhibits a similar rotational behavior, as illustrated in Fig. 3a for the representative case of $r_h = 70$ nm and $\Delta = 50$ nm. Such as strong overlap enhances the nonlinear interaction, leading to the maximization of the spatial field integral defined as $\zeta = \dfrac{\epsilon_0 \iiint \sum_{ijkl} \chi^{(3)}_{ijkl} E^*_{i,3\omega}(r) E_{j,\omega}(r) E_{k,\omega}(r) E_{l,\omega}(r) d^3r}{[\iiint \epsilon_\omega |\boldsymbol{E}_\omega(r)|^2 d^3\boldsymbol{r}]^{\frac{3}{2}} \times [\iiint \epsilon_{3\omega} |\boldsymbol{E}_{3\omega}(r)|^2 d^3\boldsymbol{r}]^{\frac{1}{2}}}$ as shown in the right panel of Fig. 3a. Note that the calculated spectra peak only at the close vicinity of the spectral position corresponding to the excitation of the QTM resonant mode (see Supplementary Note 4 for details on the derivation of the spatial mode overlap expression and more results on the spectral overlap). The measured spectral dependency of the TH signal for FF beam changing from 1400–1600 nm and fixed intensity of $\approx 1 \text{ GW/cm}^2$ (see **Methods**) is shown in Figs. 3b and 3c. Here, the air-hole radii were changed from $r_h = 90$ nm to $r_h = 130$ nm while the off-center distance was fixed to $\Delta = 50$ nm and $\Delta = 100$ nm. It is evident that as the FF beam approaches the wavelength of the nonlocal resonant mode for each case (black shaded area), the intensity of the generated TH increases and then decreases past the resonant wavelength, as shown by the different colors in each curve.

We next measure the TH signal when the FF beam is tuned to the spectral position corresponding to the nonlocal resonant mode of each sample, as shown in Fig. 3d. While satisfying the resonant condition at the FF, the designed meta-atoms not only enable the enhancement of the local field but also facilitate maximal spatial overlap between the FF and TH field distributions, leading to 1000 times enhanced TH compared to an unstructured film of the same thickness. We note that the relative magnitudes of the TH, compared to the thin film, are sensitive not only to the spectral tuning of the pump with respect to the resonance wavelength but also to material dispersion and the polarization state of the incident beam. Figure 3e shows that the normalized TH signal depends strongly on the FF wave polarization with respect to the meta-atom, where 0° corresponds to the polarization orthogonal to the off-center axis of the meta-atoms, while 90° corresponds to polarization parallel to the air-hole direction. The fitted curves in Fig. 3e, shown by the shaded regions, indicate that the collected TH signals possess an $I_{\text{THG}} \propto \cos^6(\theta)$ spatial distribution, which agrees well with the linear optical response of the arrays shown in Fig. 2, where the samples are only in resonance for x-polarization.

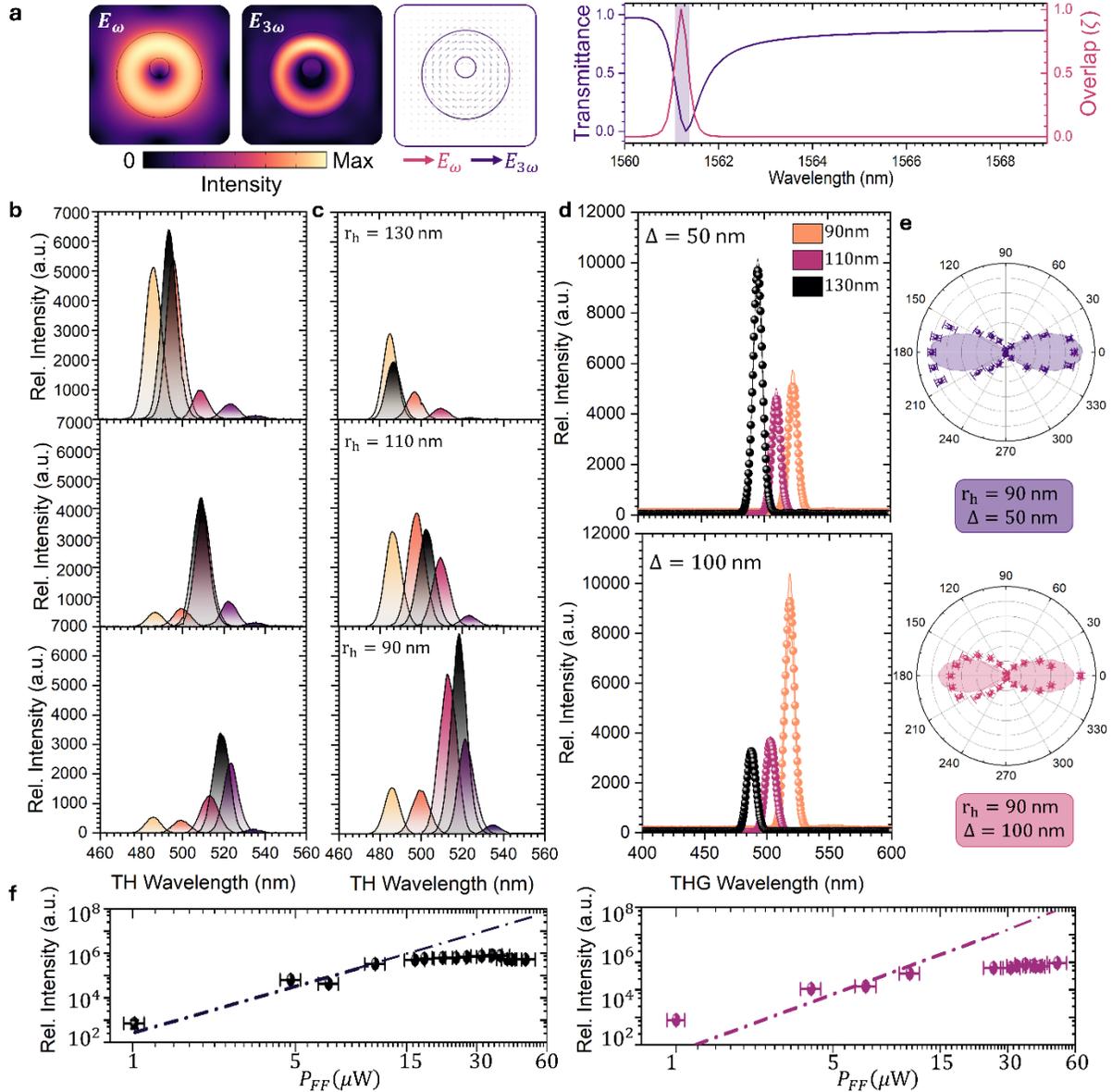

**Fig. 3 | Nonlinear response of the fabricated nonlocal metasurfaces. a,** Calculated field distributions at FF and TH wavelengths, showing maximal overlap, as also indicated by the calculated spatial mode overlap. Measured THG spectra for arrays with fixed off-center distances of **b**, $\Delta = 50$ nm and **c**, $\Delta = 100$ nm while their air-hole radii varying from $r_h = 90$ nm to $r_h = 130$ nm excited by a pump beam tuned from 1400–1600 nm. The shaded region indicates the spectral position of the nonlocal resonance. Enhanced THG is observed near resonance, with intensity decreasing away from it. **d**, THG enhancement at exact resonance wavelength, demonstrating approximately a 1000-fold increase due to optimized local-field enhancement and maximal overlap between fundamental and third-harmonic fields. **e**, Polarization-dependent THG normalized to the maximum value, revealing a $\cos^6(\theta)$ angular distribution. Here, 0° corresponds to polarization orthogonal to the meta-atoms' off-center axis, while 90° corresponds to polarization parallel to the air-hole axis. **f**, Power dependence of THG integrated photon counts for arrays with $r_h = 110, 130$ nm and $\Delta = 50$ nm, exhibiting cubic scaling at lower powers and saturation at higher intensities due to nonlinear Kerr-induced refractive index modulation and saturable nonlinearities.

We further investigated the power dependence of TH at the resonant wavelengths of arrays with $r_h = [110, 130]$ nm and $\Delta = 50$ nm, by integrating the total photon counts for the measured spectra, $c_{\text{Tot}} = \int N(\lambda) d\lambda$, as shown in Fig. 3f. At low incident pump powers, the TH signal follows a cubic power law, consistent with cubic nonlinearity, whereas, at higher pump powers, the TH response saturates and deviates from the cubic behavior. Notably, when the same high-power pump is detuned from resonance, the TH signal maintains its cubic dependence as expected due to the reduced strength of nonlinear interaction in the absence of resonant enhancement.

**Nonlocal control of diffracted orders**

Current geometric-phase-based metasurfaces predominantly rely on local interactions, leading to inherently broadband responses, whereas narrowband operation typically demands nonlocal interactions between neighboring meta-atoms, complicating precise spatial wavefront control at a specific wavelength [27-32]. However, many emerging applications, including narrowband beam steering for augmented/virtual reality (AR/VR) systems [8], [11] and nonlinear wavefront shaping for photon-pair generation [38], [39], require precise wavefront control at a specific, well-defined wavelength. For instance, in spontaneous parametric down-conversion (SPDC), photon pairs are generated over a broad spectrum; yet, only those generated at the resonant wavelength benefit from enhanced nonlinear efficiency [33-38]. In this perspective, a platform that enables geometric-phase manipulation exclusively at the resonant wavelength would allow selective control over resonantly generated photons while leaving off-resonant counterparts unaffected, thereby opening new opportunities for wavelength-specific beam shaping in quantum optics. In this section, we overcome these limitations by integrating the meta-atoms supporting nonlocal QTM resonant modes to realize simultaneous pixel-level geometric phase control and ultra-narrow spectral bandwidths. Note that increasing the spatial density of resonators (decreasing the periodicity) is vital for encoding finer spatial phase gradients, necessary for high-resolution wavefront manipulation in applications such as nonlinear holography [55-60]. However, such a reduction in periodicity leads to a shift in the resonant wavelength of the desired mode and the degradation of the achievable Q-factor. To further study this trade-off between spatial resolution and resonance fidelity, we investigated the linear optical response of a metasurface with $r_h = 70$ nm and $\Delta = 50$ nm as a function of its lattice periodicity changing from 650 nm to 950 nm and wavelength. **Figure 4a** shows the simulated and measured spectra indicating the presence of three resonance branches for each lattice periodicity. Two of these resonances occur at shorter wavelengths, while the third resonance emerges at a longer wavelength. Specifically, the shorter-wavelength resonances exhibit local characteristics (see Supplementary Note 5 for further details) and gradually merge into a single resonant mode at larger periodicities with its electric and magnetic field distributions shown in Fig. 4b. In contrast, larger lattice periodicity leads to the resonant wavelength of the third resonance branch shift from 1490 to 1650 nm, while its spectral linewidth remains unchanged, as it is primarily determined by the size and off-center positioning of the hole within each nanodisk. Therefore, this resonance is robust against changes in lattice period and gradually shifts toward longer wavelengths as the periodicity increases. The corresponding field distribution, shown in Fig. 4b, together with the measured full width at half maximum (FWHM), illustrated in Fig. 4c, further confirms that this resonance corresponds to the nonlocal mode, with its Q-factor being robust against variations in lattice periodicity.

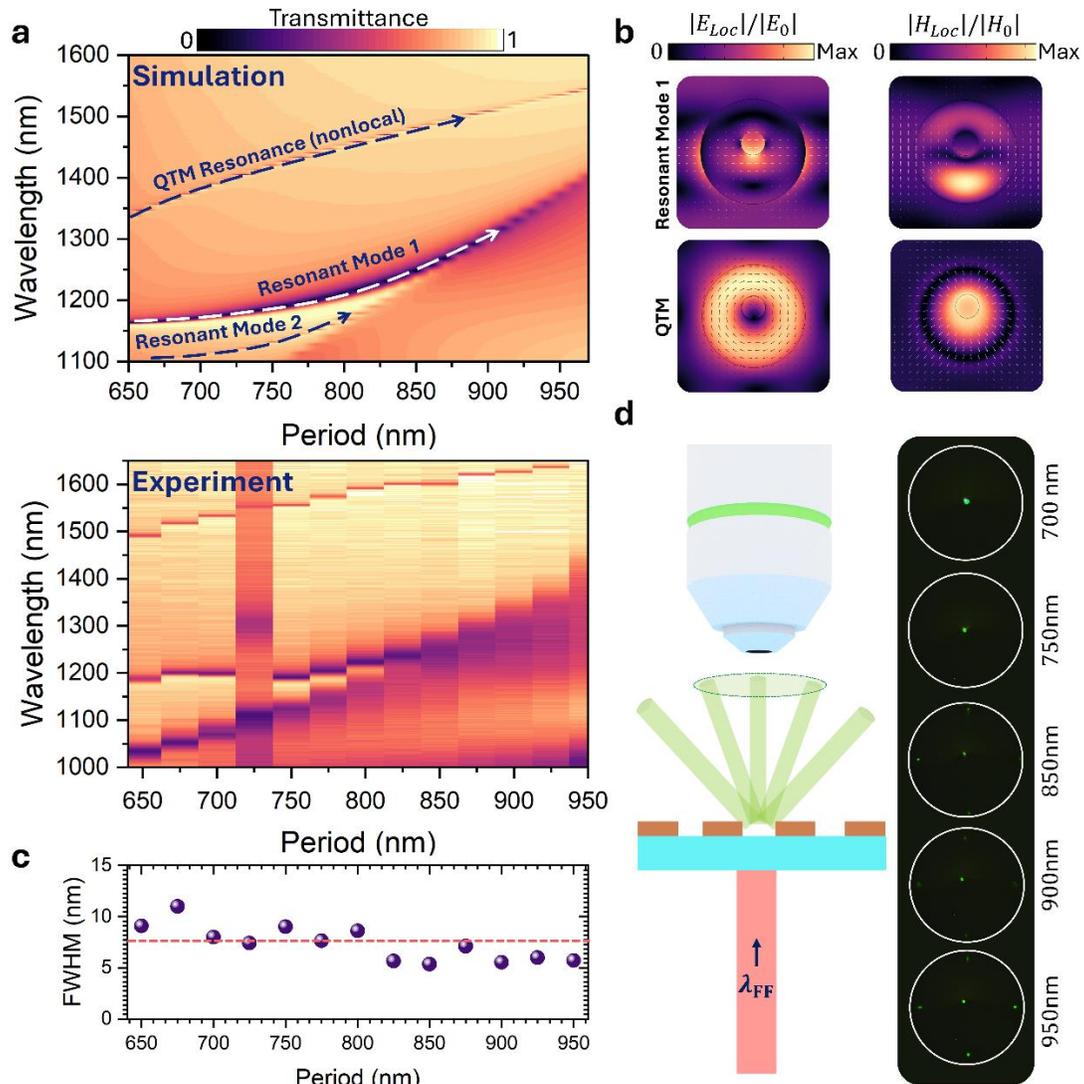

**Fig. 4| Spectral evolution and diffraction behavior of the fabricated nonlocal metasurfaces. a**, Simulated and measured transmission spectra of metasurfaces with varying lattice periodicities (650–950 nm) reveal three distinct resonance branches. Two resonances at shorter wavelengths exhibit local characteristics and merge at larger periodicities, while a third, narrowband resonance shifts toward longer wavelengths with increasing periodicity. **b**, Electric and magnetic field distributions of the resonances for the fixed periodicity of $P = 900$ nm. **c**, Measured FWHM of the emerged nonlocal resonant mode as function of metasurface lattice periodicity, indicating nearly constant behavior across different periodicities. The dashed line denotes the average value of FWHM. **d**, Schematic illustration of third-harmonic diffraction behavior (left) and measured diffraction spots (right). For periodicities below 825 nm, only the zeroth-order is collected due to limitations in collection optics, whereas for larger periodicities, higher diffraction orders are captured, resulting in enhanced THG efficiency.

Despite the robustness of the designed nonlocal resonant mode at the FF, Figure 4d shows that the third-harmonic signal is distributed among multiple diffraction orders due to the onset of diffraction effects when the wavelength becomes comparable to the lattice constant, as in the case at TH. In particular, the intensity distribution among the diffracted orders results from a complex interplay between the spatial profile of the resonantly enhanced fields and the geometry of the meta-atoms. However, due to limitations of the collection optics in our experimental setup, only a subset of these diffracted orders can be captured, as also indicated in Fig. 4d. Specifically, for lattice constants below 825 nm, only the zeroth diffraction order is collected (See Supplementary Note 6 for more results), while as the periodicity increases beyond this threshold, higher diffraction orders fall within the collection angle, leading to an increase in the experimentally measured TH generation efficiency. Next, we focus on two representative metasurfaces with lattice constants of 750 nm and 1000 nm, respectively, and demonstrate how high Q-factor nonlinear nonlocal metasurfaces can lead to wavefront manipulation.

**Nonlocal nonlinear wavefront manipulation**

As discussed earlier, the introduction of the off-center hole into the meta-atom changes its $C_\infty$ rotational symmetry to $C_1$, thereby enabling spin–orbit coupling in the nonlinear regime [19]. In particular, conventional selection rules indicate that for a circularly polarized light beam illuminating an arbitrarily rotated meta-atom, the allowed Berry phases can be expressed as $\phi_{Co} = (n-1)\sigma\alpha$ and $\phi_{Cross} = (n+1)\sigma\alpha$, with $\sigma = \pm 1$ representing the helicity of the incoming beam, $\alpha$ is the in-plane rotation of the meta-atom along the $z$-axis, and $n$ denoting the harmonic order with $n = 3$ indicating THG. Therefore, the angular displacement of the hole within each meta-atom imparts a geometric phase of $2\alpha$ to the cross-polarized component at the FF, and phase shifts of $2\alpha$ and $4\alpha$ to the co- and cross-polarized third-harmonic signals, respectively. In general, when a circularly polarized light passes through the metasurface, the polarization conversion efficiency is given by $\eta = \left|t_x - t_y\right|^2/2$, where $t_x$ and $t_y$ denote the complex transmission coefficients for $x-$ and $y$-polarized light, respectively. As recently discussed in [61], for a metasurface with different responses under $x-$ and $y$-polarization, the conversion efficiency modifies to $\eta = \frac{1}{4}\gamma^2/[(\omega-\omega_0)^2 + \gamma^2]$, with $\omega_0 + i\gamma$ representing the resonant frequency and its associated damping factor. In our case, the metasurface possesses a strong nonlocal resonant response under x-polarized illumination and remains nonresonant under y-polarized excitation. Therefore, at the spectral position corresponding to the quasi-trapped mode resonance ($\omega = \omega_{QTM} = \omega_0$), the maximum polarization conversion transmittance is theoretically expected to reach $\eta|_{\omega=\omega_0} = \frac{1}{4}\gamma^2/[(\omega_0-\omega_0)^2 + \gamma^2] = 0.25$. We note that although the metasurface design was primarily optimized for linearly polarized light, any arbitrary polarization state, such as circular polarization, can be decomposed into orthogonal linear components. As a result, the metasurface maintains its nonlocal resonant behavior under CP illumination, as shown in **Fig. 5a** for metasurfaces with lattice periodicity $P = 1000$ nm and asymmetry parameters $(r_h, \Delta) = (130,50)$ and $(90,100)$ nm (See Supplementary Note 7 for results corresponding to $P = 750$nm).

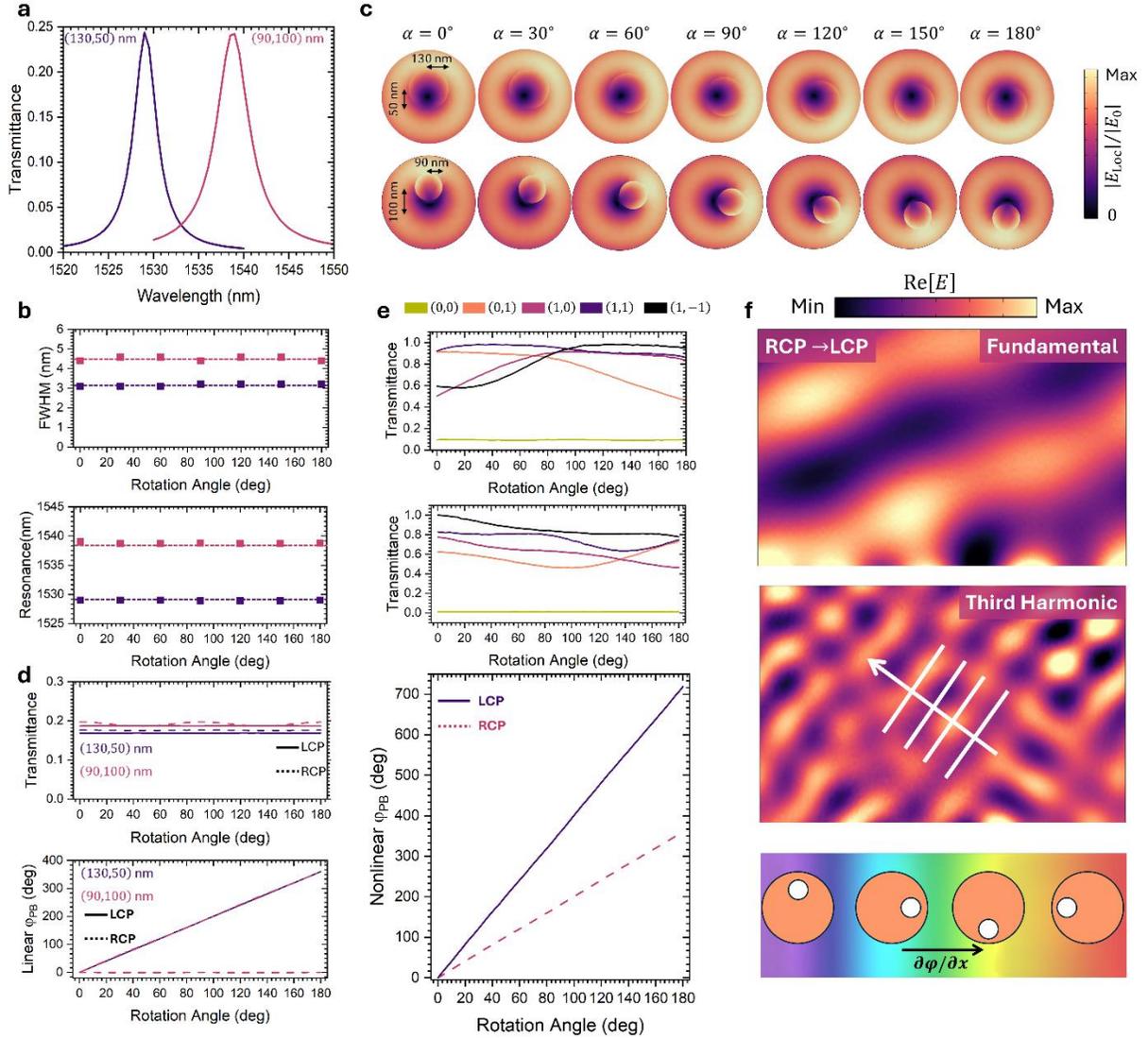

**Fig. 5| Pixel-level linear and nonlinear wavefront control enabled by high-Q nonlocal metasurfaces. a**, Measured transmittance under RCP illumination for metasurfaces with $P = 1000$ nm and asymmetric geometries $(r_h, \Delta) = (130, 50)$ and $(90, 100)$ nm. A maximum polarization conversion efficiency of 0.25 is achieved at the QTM resonance, confirming operation exclusively at the designed frequency. **b**, Resonant spectra as a function of meta-atom rotation angle changing from $0 < \alpha < \pi$, showing that the resonance wavelength remains invariant under rotation, in contrast to conventional nonlocal metasurfaces. **c**, Electric field distributions in the $(x - y)$ plane for different rotation angles further confirm that the nonlocal character of the mode remains unchanged. **d**, Retrieved amplitude and phase of the fundamental transmitted light under RCP illumination. Only the zeroth diffraction order is present, with the cross-polarized (LCP) component acquiring a geometric phase of $\varphi_{PB} \approx 2\alpha$, while the co-polarized (RCP) component remains constant in phase. **e**, Nonlinear (THG) co- and cross-polarized amplitude and phase as functions of rotation angle. The (0,0) order shows minimal amplitude variation and follows $2\alpha$ and $4\alpha$ phase dependence for RCP and LCP, respectively, while higher diffraction orders exhibit stronger amplitude modulation (~60%). **f**, Top: Linear beam steering at the FF wavelength using four meta-atoms with 360° geometric phase coverage with high Q-factor. Bottom: Nonlinear beam steering under the same design exhibits a distinct deflection angle due to a different nonlinear phase gradient. Fluctuations in the nonlinear wavefront arise from interference with higher diffraction orders and can be mitigated by structural optimization to favor the (0,0) order.

**Figure 5a** confirms that under CP excitation, the maximum transmittance reaches the expected value of 0.25 at resonance and drops to zero off-resonance, confirming that the metasurface operates exclusively at the designed resonant frequency. To demonstrate that the imparted geometric phase can be controlled on a meta-atom level at a resonant wavelength, we first continuously rotate the nanoresonators over the range $0 < \alpha < \pi$ and record their corresponding resonant spectra, as shown in Fig. 5b. As confirmed in this figure, the spectral position of the resonant mode remains unchanged across all rotation angles for both structures, which highlights a key advantage of our design compared to conventional nonlocal metasurfaces with $p2$ symmetry, where the resonant wavelength typically shifts as a function of the rotation angle [12-14]. The electric field distribution in the $(x - y)$ plane as a function of rotation angle, shown in Fig. 5c, further confirms that the overall behavior of the field profiles remains largely unchanged with varying rotation angle, with the primary difference being a shift in the position of the maximum electric field intensity.

To demonstrate meta-atom level phase control of both the FF and TH, we evaluate the co- and cross-polarized amplitude and phase responses as a function of meta-atom rotation angle under right-handed circularly polarized illumination, as shown in Fig. 5d. At the FF, where the structure is subwavelength in scale, only the zeroth-order (0,0) diffraction mode exists, and the amplitudes of the transmitted right circularly polarized (RCP, dashed line) and left circularly polarized (LCP, solid line) components remain nearly constant with respect to the rotation angle. In contrast, the retrieved phase profiles reveal that, at the resonant wavelength, the cross-polarized (LCP) component acquires a geometric phase of approximately $\varphi_{PB} \approx 360°$ (For the rotation angle of $\alpha = 180°$) in excellent agreement with the theoretical prediction $\varphi_{\text{cross}} = 2\alpha$, whereas the co-polarized component shows no geometric phase accumulation. These results indicate that a single metasurface can simultaneously manipulate the wavefront at the meta-atom level with a high Q-factor and engineer the spectral response. To illustrate this capability, we present an example of beam steering at the resonant wavelength, as shown in Fig. 5f (top). According to the generalized Snell's law [62], the deflection angle $\theta_{FF}$ is related to the applied phase gradient $\partial \varphi / \partial x$ by $\sin(\theta_{\text{FF}}) = (\lambda_{FF}/2\pi) \, \partial \varphi / \partial x + m\lambda_{FF}/P$, where $m$ denotes the diffraction order and $P$ is the lattice periodicity. As a simple yet illustrative example, we implement a full 360° geometric phase coverage using only four meta-atoms, resulting in a deflection of cross-polarized components of the zeroth-order spot at the FF while maintaining a high Q-factor, as shown in Fig. 5f. We note that the co-polarized component of the transmitted wave remains undeflected due to the absence of any phase accumulation, as shown in Supplementary Note 8.

We next demonstrate that our high-Q PB-based approach to wavefront manipulation can also be extended to control the phase of generated nonlinear harmonics at the nanoscale, on a pixel-by-pixel basis. Figure 5e shows the normalized amplitude and phase responses of the co- and cross-polarized TH signals as a function of meta-atom rotation angle under right-handed circularly polarized excitation. At the TH wavelength, the metasurface periodicity exceeds the harmonic wavelength, resulting in multiple diffraction orders for both RCP and LCP components, as indicated by different colors. We note that the field distribution within the nanoresonators directly influences the amplitude ratio between these polarization states. In particular, the (0,0) nonlinear diffraction order shows minimal amplitude variation with respect to the meta-atom rotation angle, indicating that amplitude modulation for this order is negligible, which may be a desirable feature for holographic display applications [55-60]. In contrast, while the higher-order diffraction components exhibit greater efficiency than the (0,0) order, they display amplitude variations of approximately 60% as a function of the rotation angle. Therefore, a trade-off emerges between achieving higher conversion efficiency at the expense of wavefront uniformity and maintaining minimal amplitude variations at the cost of reduced overall efficiency. On the other hand, the imparted nonlinear geometric phase on the (0,0) third-harmonic components follows the expected behavior: the RCP (dashed line) and LCP (solid line) signals acquire phase shifts of $2\alpha$ and $4\alpha$, respectively, while, for higher-order diffraction modes, the nonlinear phase behaves differently due to helicity mixing. In particular, such mixing arises because, when the fixed laboratory circular polarization basis is applied to diffraction lobes whose true circular polarization vectors lie in planes perpendicular to their tilted wavevectors, the intrinsic nonlinear Pancharatnam–Berry phases are projected into a geometry-dependent weighted superposition (see Supplementary Note 8 for results on additional higher diffraction orders).

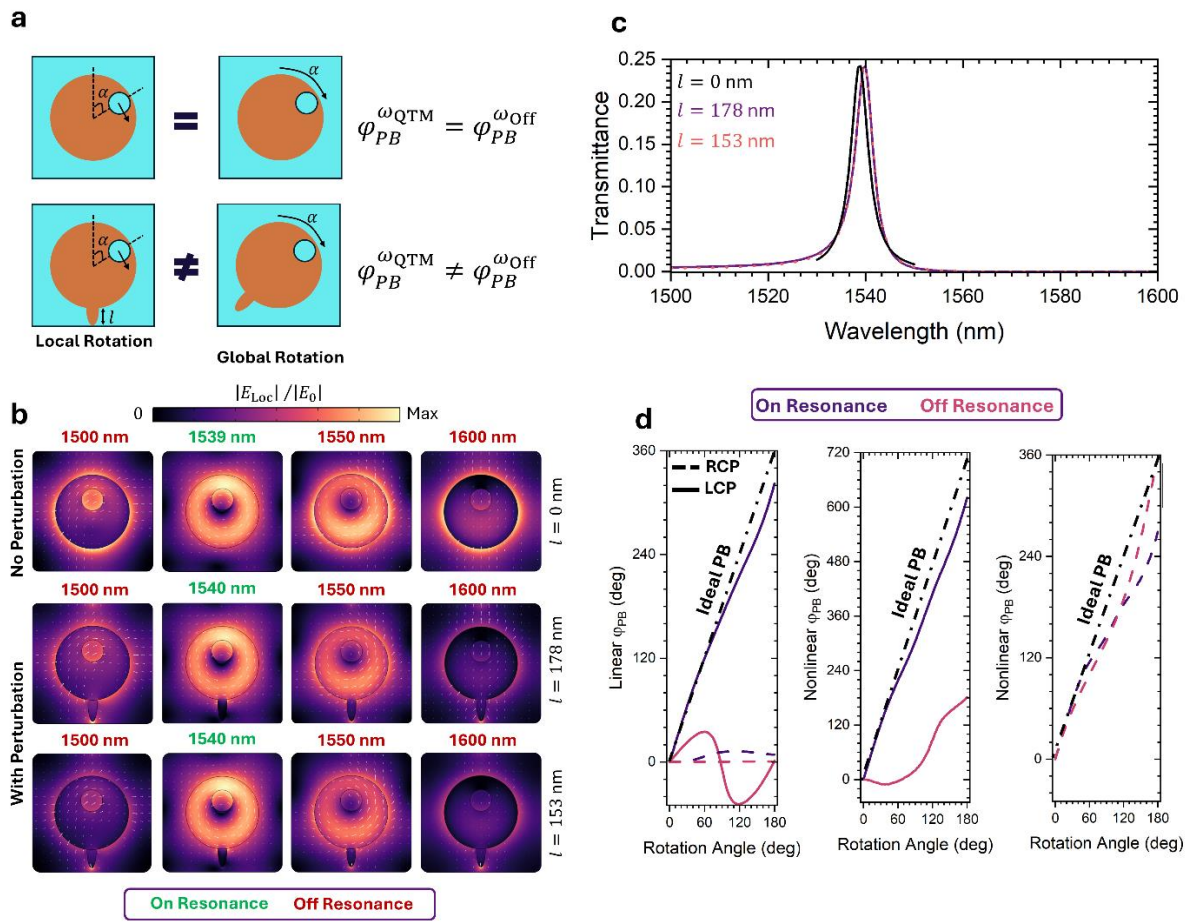

**Fig. 6| Resonant-wavelength-selective geometric phase control enabled by perturbation-engineered nonlocal metasurfaces.**
**a**, Schematic comparison of local and global rotations. In the absence of perturbation (top), local and global rotations are equivalent and yield conventional PB phases. With perturbation (bottom), local rotation no longer corresponds to a global rotation, enabling wavelength-selective geometric phase control. **b**, Simulated electric field distributions at resonant and off-resonant wavelengths for unperturbed (top) and perturbed (bottom) meta-atoms. At resonance, fields are confined within the nanoscale holes, whereas off-resonant fields are distributed near the outer boundaries. The introduced half-ellipsoidal perturbation alters the off-resonant field distribution while preserving the resonant mode. **c**, Transmission spectra showing that the perturbation has minimal effect on the resonant wavelength while significantly modifying the optical response at off-resonant wavelengths. **d**, Unwrapped phase responses under RCP illumination as a function of meta-atom rotation angle. At the fundamental frequency (left), cross-polarized components acquire the expected $2\alpha 2$ PB phase only at resonance, while off-resonant components show negligible phase variation. At the third-harmonic frequency, the co-polarized component (middle) follows $2\alpha 2$, whereas the cross-polarized component (right) exhibits conventional $4\alpha$ behavior at resonance and a newly observed linear dependence at off-resonance. This resonant-only phase tunability enables independent control of geometric phase at FF and TH wavelengths, allowing for selective wavefront manipulation and the generation of structured beams exclusively at the resonant frequency.

We have also demonstrated nonlinear wavefront manipulation of the same structure (four meta-atoms covering full 360°), and the results are shown in Fig. 5f (bottom). As observed, the deflection angle of the third-harmonic signal differs from that of the fundamental beam, which is attributed to the different phase gradient that is imparted on the nonlinear wavefront (i.e., $\varphi_{+\sigma}^{(3\omega)} = 2\alpha$ and $\varphi_{-\sigma}^{(3\omega)} = 4\alpha$) compared to the FF beam ($\varphi_{+\sigma}^{(\omega)} = 0$ and $\varphi_{-\sigma}^{(\omega)} = 2\alpha$). The presence of higher diffraction orders and their interference with the (0,0) order, leading to fluctuations in the nonlinear wavefront, which can be further suppressed by modifying the structure to favor only the zeroth-order diffraction while maintaining resonant conditions. We note that while we have thus far demonstrated both linear and nonlinear wavefront steering using our proposed nonlocal high-Q resonant metasurface, which, to the best of our knowledge, represents the first demonstration of such functionality in the nonlinear regime, the imparted geometric phase spans a broad spectral range. This broadband nature limits precise phase control to the resonant wavelength, which is critical for narrowband beam steering in AR/VR systems [8,11] and for nonlinear wavefront shaping in photon-pair generation [38,39]. We overcome this limitation by slightly modifying the meta-atom geometry through the addition of a nanoscale elliptical perturbation, as shown in Fig. 1d. Such a small perturbation confines the imparted geometric phase to the resonant wavelength in both the linear and nonlinear regimes, rendering the off-resonant response insensitive to meta-atom rotation and giving rise to entirely new nonlinear Pancharatnam–Berry selection rules in the third-harmonic spectrum.

Here, we highlight an important distinction between the two types of rotational transformations, namely, local and global transformations, as shown in **Fig. 6a**. The local rotation involves rotating the center of the nanoscale hole according to the parametric equations $x = \Delta \sin(\alpha)$ and $y = \Delta \cos(\alpha)$ around the axis of the cylinder. This transformation is equivalent to a global rotation, where the entire structure, including both the cylinder and the hole, is rotated by an angle $\alpha$ around the $z$-axis, as shown in the top panel of Fig. 6a. Both local and global rotations correspond to the conventional PB phase in the linear and nonlinear regimes, leading to predictable polarization-dependent phase shifts. Specifically, for a circularly polarized input with spin $\sigma$, the linear PB phase follows $\varphi_{+\sigma}^{(\omega)} = 0$ and $\varphi_{-\sigma}^{(\omega)} = 2\alpha$, while the nonlinear PB phase at the third-harmonic frequency follows $\varphi_{+\sigma}^{(3\omega)} = 2\alpha$ and $\varphi_{-\sigma}^{(3\omega)} = 4\alpha$, as previously demonstrated in Fig. 5. We note that these imparted geometric phases are independent of wavelength, resulting in identical phase shifts at both resonant and nonresonant wavelengths, which in turn, prevents selective wavefront shaping confined to the resonant wavelength. However, when a judiciously designed perturbation in the form of a half-ellipsoidal with length $l$ is introduced at the boundary of the meta-atoms, the rotation of the hole (local rotation) no longer corresponds to a global rotation of the entire structure, as illustrated in the bottom panel of Fig. 6b. This structural distinction becomes physically meaningful when considering the markedly different spatial distributions of the optical modes at on- and off-resonance. In particular, for the unperturbed structure, the field associated with the QTM resonance is predominantly confined to the interior of the nanostructure (especially near the air-hole). In contrast, the nonresonant fields are distributed uniformly near the outer boundary, as shown in the first row of Fig. 6b. When the perturbation is added, the field profile of the resonant mode remains largely unchanged with respect to hole rotation, while at off-resonant wavelengths, the optical field overlaps primarily with the outer regions of the structure, where the perturbation remains fixed as shown in bottom rows of Fig. 6b for $l = 178$ nm and $l = 153$ nm (see Supplementary Note 9 for more details on the effect of rotation). In this perspective, the effective symmetry experienced by the nonresonant modes, that is, the symmetry of the structure as sampled by the spatial extent of the field distributions, does not vary with hole rotation. As a result, while the far-field polarization and phase of the resonant mode vary with the hole angle $\alpha$, the nonresonant response remains nearly invariant, decoupling the PB phase at the resonant wavelength from that at off-resonant wavelengths. We note that such a boundary perturbation must be carefully designed to satisfy the following criteria: (i) it must be small enough to preserve the mode distribution at the resonant wavelength, maintaining both rotational symmetry and field confinement around the holes; (ii) it must not introduce a significant shift in the resonant wavelength, thereby preserving the nonlocal integrity of the design; and (iii) it must be large enough to produce a non-negligible effect on the optical field at off-resonant wavelengths, ensuring that the effective symmetry experienced by the nonresonant modes remains insensitive to hole rotation. Based on these design principles, the representative cases of $l = 178$nm and $l = 153$ nm were selected, as evidenced by their corresponding field distributions at both resonant and off-resonant wavelengths and their associated transmittance spectra shown in Fig. 6c (see Supplementary Note 10 for results on other perturbation dimensions). As a consequence of the perturbation-induced distinction between resonant and nonresonant field distributions, the geometric phase imparted at each wavelength can be defined independently, ultimately leading to distinct sets of geometric phases, as has also been very recently discussed in the linear regime [63]. In this perspective, the nonlinear geometric phases are also expected to be modified, given their correlation with the linear geometric phase behavior.

To demonstrate phase control at both the FF and TH generation wavelengths at the meta-atom level, we evaluate the unwrapped co- and cross-polarized phase responses of the perturbed structure as a function of meta-atom rotation angle under right-handed circularly polarized illumination, as shown in Fig. 6d. At the FF wavelength (left panel), the retrieved phase profiles reveal that at resonance, the LCP component acquires a geometric phase of $\varphi_{-\sigma}^{(\omega)} \approx 2\alpha$. In contrast, the co-polarized component exhibits no phase modulation, consistent with conventional PB behavior. Interestingly, at an off-resonant wavelength (e.g., 1500 nm), the cross-polarized component no longer follows the expected $\varphi_{-\sigma}^{(\omega)} = 2\alpha$ dependence, instead showing minimal variation with respect to the rotation angle, similar to the co-polarized component, which also exhibits no phase response. This is a direct consequence of the selective response between resonant and nonresonant modes, which arises from fundamental differences in spatial symmetry and field localization. At resonance, the QTM mode is strongly confined near the center of the meta-atom, making the system sensitive to local rotation and allowing the imparted phase to follow the orientation of the meta-atom. In contrast, off-resonant fields are primarily distributed near the outer boundary, where the added perturbation remains fixed. As a result, the symmetry experienced by the off-resonant modes does not vary with rotation, and no geometric phase is imparted. On account of such a decoupled behavior in the linear spectrum, the corresponding nonlinear geometric phase is also modified at both resonant and nonresonant wavelengths, as shown for the co-polarized (middle panel) and cross-polarized (right panel) components of the TH signal in Fig. 6d.

As shown in the middle panel of Fig. 6d, the LCP component at the resonant wavelength exhibits a nonlinear geometric phase that closely follows the ideal nonlinear Pancharatnam–Berry phase expected for the cross-polarized component. In contrast, at the off-resonant wavelength, the phase behavior changes entirely, exhibiting a new selection rule that follows a nonlinear dependence on the meta-atom orientation—a phenomenon that, to the best of our knowledge, is observed here for the first time. This unconventional nonlinear geometric phase arises directly from the added perturbation, which alters the off-resonant field distribution at the meta-atom boundary, effectively decoupling the global and local rotational symmetries of the structure. For the co-polarized component, the nonlinear geometric phase remains consistent with the conventional PB phase, following $\varphi_{+\sigma}^{(3\omega)} \approx 2\alpha$. This behavior is expected, as the co-polarized fundamental component is also unaffected by the perturbation, and the nonlinear PB phase is inherently linked to its linear counterpart. Consequently, the linear and nonlinear geometric phases in the resonant nonlocal metasurface case are jointly governed by the meta-atom geometry and internal field distribution, allowing for independent control of phase at both resonant and nonresonant wavelengths. Based on this design, diverse third-harmonic wavefronts for both RCP and LCP can be achieved at the resonant wavelength, leaving the off-resonant spectrum unaffected. The unique combination of resonant-only geometric phase manipulation and strong field enhancement provides a versatile platform for nonlinear, nonlocal wavefront engineering, opening up new pathways for generating complex, structured beams at harmonic frequencies.

**Discussion**

While metasurfaces offer a promising alternative to bulk nonlinear crystals, existing designs face a trade-off between high nonlinear efficiency and precise wavefront control due to the conflicting nature of local and nonlocal responses. Here, we present a nonlinear nonlocal metasurface that supports quasi-trapped resonant modes, enabling both efficient harmonic generation and pixel-level phase manipulation. We theoretically and experimentally demonstrate that our high-Q metasurface, with a measured Q-factor of $Q \approx 500$, maintains global symmetry while allowing local parameter variation, thereby achieving continuous geometric phase control at both the fundamental and third-harmonic wavelengths. The engineered field distribution enhances spatial mode overlap between the FF and TH fields, resulting in a three orders of magnitude increase in TH efficiency compared to unstructured thin films. As proof of concept, we demonstrated beam steering of both the fundamental and third-harmonic beams using a metasurface composed of only four distinct meta-atoms. We further demonstrated that introducing a carefully designed perturbation to the meta-atom geometry results in geometric phase accumulation at the resonant wavelength, but not off resonance—a behavior not observed in previously reported linear and nonlinear PB-based metasurfaces. This selectivity originates from the unique QTM field profile, which preserves global symmetry off-resonance while enabling local geometric phase at resonance for both the FF and TH signals. To our knowledge, this is the first demonstration of simultaneous nonlinear generation and wavefront manipulation at resonance in a nonlocal metasurface. We believe these results not only expand the functional scope of silicon photonics for advanced optical communication and quantum information systems but also uncover new physical mechanisms in nonlinear geometric phase control at the nanoscale.

## Methods

### Samples preparation

A 100 nm amorphous silicon film was deposited on the glass substrate using PECVD. Afterwards, a 100nm polymethyl methacrylate (PMMA-A2) film was spin-coated onto the deposited film and baked at 180 °C for two minutes. The desired meta-atom shapes were transferred to the PMMA resist using electron beam lithography (EBL) (Elionix ELS-7500 EX) and developed in a MIBK/IPA solution for 60 seconds at room temperature. The sample was transferred into an e-beam evaporator and directly coated with 30 nm of chromium (Cr) film at a deposition rate of 5 Å/s. After immersing the sample in Remover 1165 for 12 hours, the PMMA was removed, and the nanostructures were transferred to Cr. The a-Si under the Cr-covered nanostructure was then etched by an Inductively Coupled Plasma (ICP)-RIE etching (Oxford PlasmaPro ICP Etcher), and the remaining Cr film was removed by immersing the sample into the chromium etchant for 10 minutes (See Supplementary Information for more details).

### Numerical Simulations

The numerical simulations were conducted using the finite element method (FEM) implemented in the commercial software COMSOL Multiphysics. In particular, we utilized the Wave Optics Module to solve Maxwell's equations in the frequency domain, incorporating proper boundary conditions. The meta-atoms were studied under the plane wave illumination along the $z$-axis with an electric field pointing along the $x$-axis. We applied periodic boundary conditions in the x and y directions and the PML in the z direction to avoid undesired reflections. Periodic ports were used along the z direction to launch a plane wave and capture the transmitted power. For the nonlinear simulations, we utilized the undepleted pump approximation. This involved a two-step process to compute the intensity of the radiated nonlinear signal. Initially, linear scattering at the pump wavelength was simulated to derive the induced nonlinear polarization within the meta-atom. This induced polarization was then used as a source in the electromagnetic simulation at the harmonic wavelength to determine the TH field produced. The far-field phase profiles at the fundamental and third-harmonic wavelengths are retrieved using a custom MATLAB code based on near-field to far-field transformations.

For the isolated meta-atom case, we used a spherical air-filled domain and a radius of $4\lambda$ as the background medium. At the same time, perfectly matched layers of thickness $0.6\lambda$ were positioned outside the background medium to act as absorbers and prevent backscattering. The tetrahedral mesh was chosen to ensure the accuracy of the results and facilitate numerical convergence. Upon the interaction of light with the meta-atom, the induced polarization is related to the field distributions within the particle via $\boldsymbol{P} = \epsilon_0(\epsilon_p - \epsilon_d)\boldsymbol{E}_p$, where $\epsilon_0$, $\epsilon_p$, and $\epsilon_d$ are the free space, particle, and surrounding medium dielectric constants, respectively, and $\boldsymbol{E}_p$ is the total electric field inside the scatterer. The scattered field is given by the superposition of different multipole moments (up to the electric octupole term) as [76]

$$\boldsymbol{E}_{sct}(\boldsymbol{n}) = \frac{k_0^2 \exp(ik_0 r)}{4\pi\epsilon_0 r}\left([\boldsymbol{n} \times [\boldsymbol{D} \times \boldsymbol{n}]] + \frac{1}{c}[\boldsymbol{m} \times \boldsymbol{n}] + \frac{ik_0}{6}[\boldsymbol{n} \times [\boldsymbol{n} \times \hat{Q}\boldsymbol{n}]] + \frac{ik_0}{2c}[\boldsymbol{n} \times \widehat{M}\boldsymbol{n}] \right. \\ \left. + \frac{k_0^2}{6}[\boldsymbol{n} \times [\boldsymbol{n} \times \hat{O}(\boldsymbol{nn})]]\right), \quad (2)$$

where $\boldsymbol{D}$ corresponds to the total electric dipole (ED) moments, $\boldsymbol{m}$ is the exact magnetic dipole (MD) moment, and $\hat{Q}$, $\hat{O}$, and $\widehat{M}$ represent the electric quadrupole (EQ), electric octupole (EO), and magnetic quadrupole tensors, respectively; $\boldsymbol{n} = \boldsymbol{r}/r$ is the unit vector directed from the particle's center towards an observation point, and $c$ is the speed of light in vacuum (See Supplementary Note 9 for details). Using these notations, the far-field scattered power can be readily related to the scattered fields of Equation (1) with the aid of the time-averaged Poynting vector as $dP_{\text{Sct}} = 0.5\sqrt{\epsilon_0/\mu_0}|E_{\text{Sct}}|^2 r^2 d\Omega$, wherein $d\Omega = \sin\theta d\theta d\varphi$ represents the solid angle []. Therefore, combining Equation (2) with the given relation of scattered power and performing the integration over the total solid angle, the scattering cross-section, which is defined as $\sigma_{\text{Sct}} = P_{\text{Sct}}/P_{in}$, with $P_{in}$ being the incident power.

**Linear Optical Characterization**

A stabilized fiber-coupled light source (*Thorlab SLS201*) was used as the white light source to illuminate the meta-atoms from 300 nm to 2600 nm. A Glan-Thompson polarizer (*Thorlab GTH10*) was implemented to polarize the light beam along the desired direction, and the incident beam was weakly focused on the sample with the illumination beam wavevector distribution within the range of angles $|\theta_i| < 10°$. The transmitted light was then collected by a 10X infinity-corrected objective with the numerical aperture NA = 0.25 (*Olympus Plan Achromat Objective*) and partially guided to a camera and $100\mu m$ pinhole with a beam splitter. The setup used a pinhole to avoid undesired responses from the bare substrate and improve the signal-to-noise ratio. The transmitted light was then coupled to a large diameter (600 $\mu m$) multimode fiber via a 200mm lens (Thorlabs LB1945) and sent to a wide-range optical spectrum analyzer (*AQ6374 OSA*) to perform transmittance measurements (See Supplementary Note 10 for details of the setup). The transmittance spectra were then measured as $T(\lambda) = I(\lambda)/I_0(\lambda)$ with $I(\lambda)$ denoting the transmitted wavelength-dependent intensity arriving at the detector when the sample was inserted into the beam path, and $I_0(\lambda)$ is the transmitted wavelength-dependent intensity arriving at the detector when the reference substrate was inserted into the beam path. To ensure the repeatability of the experimental results, we performed 20 measurements for each sample and reported their average as the final response along with their standard deviation as $\sigma = \sqrt{\frac{1}{N-1}\Sigma_i(M_i - \bar{M})^2}$, with $M_i$ and $\bar{M}$ denoting each measurement and their average, respectively.

**Nonlinear Optical Characterization**

An 800 nm beam was generated from a Ti: sapphire laser (Libra system, Coherent) operating at a 1 kHz repetition rate with 100 fs pulse duration, and was directed into an ultrafast optical parametric amplifier (TOPAS-C, Light Conversion), providing tunable output across a spectral range of 260–2600 nm. A 4f system comprising two lenses was used for sample alignment and imaging the arrays. The incident beam was weakly focused onto the sample using a 75 mm $CaF_2$ plano-convex lens, and the generated TH signal was collected using an AO 40× Long Working Distance (LWD) Plan Achro objective (NA = 0.55). The residual fundamental frequency was suppressed using a set of short-pass filters. The TH photons were coupled into a multimode fiber (NA = 0.50, Ø400 μm core, FP400URT, Thorlabs) connected to a UV–VIS–NIR spectrometer (Super Gamut, BaySpec Inc.). A manual filter wheel with neutral density filters was used to vary the power of the fundamental beam. Additional details on the optical setup can be found in Supplementary Note 11.

**Acknowledgments**

This paper was supported in part by the Army Research Office (W911NF2310057), National Science Foundation (NSF) (Grant No. 2240562) and North Atlantic Treaty Organization (NATO) Science for Peace and Security program (G5984). Samples etching were performed at the Center for Nanophase Materials Sciences, which is a DOE Office of Science User Facility.

**Conflict of Interest**

The authors declare no conflict of interest.

**Data Availability Statement**

The data that support the findings of this study are available from the corresponding author upon reasonable request.

**Author contributions**

N.M.L. and H.B.S initiated the idea of this study. H.B.S, A.R.F, L.C. and M.A.V. conducted theoretical and numerical studies. The samples were fabricated by Y.Z. and H.B.S and later were etched by I.K. All authors contributed to the design and discussions. N.M.L. supervised the study performed in this work. All authors collectively contributed to the writing of the manuscript.

**References**
1. Chen, Hou-Tong, Antoinette J. Taylor, and Nanfang Yu. "A review of metasurfaces: physics and applications." Reports on progress in physics 79.7 (2016): 076401.

2. Chen, Wei Ting, Alexander Y. Zhu, and Federico Capasso. "Flat optics with dispersion-engineered metasurfaces." Nature Reviews Materials 5.8 (2020): 604-620.
3. Genevet, Patrice, et al. "Recent advances in planar optics: from plasmonic to dielectric metasurfaces." Optica 4.1 (2017): 139-152.
4. Shaltout, Amr M., Vladimir M. Shalaev, and Mark L. Brongersma. "Spatiotemporal light control with active metasurfaces." Science 364.6441 (2019): eaat3100.
5. Shastri, Kunal, and Francesco Monticone. "Nonlocal flat optics." Nature Photonics 17.1 (2023): 36-47.
6. Overvig, Adam, and Andrea Alù. "Diffractive nonlocal metasurfaces." Laser & Photonics Reviews 16.8 (2022): 2100633.
7. Kolkowski, Radoslaw, et al. "Nonlinear nonlocal metasurfaces." Applied Physics Letters 122.16 (2023).
8. Song, Jung-Hwan, et al. "Non-local metasurfaces for spectrally decoupled wavefront manipulation and eye tracking." Nature Nanotechnology 16.11 (2021): 1224-1230.
9. Hail, Claudio U., et al. "High quality factor metasurfaces for two-dimensional wavefront manipulation." Nature Communications 14.1 (2023): 8476.
10. Klopfer, Elissa, et al. "High-quality-factor silicon-on-lithium niobate metasurfaces for electro-optically reconfigurable wavefront shaping." Nano letters 22.4 (2022): 1703-1709.
11. Lawrence, Mark, et al. "High quality factor phase gradient metasurfaces." Nature Nanotechnology 15.11 (2020): 956-961.
12. Malek, Stephanie C., et al. "Multifunctional resonant wavefront-shaping meta-optics based on multilayer and multi-perturbation nonlocal metasurfaces." Light: Science & Applications 11.1 (2022): 246.
13. Malek, Stephanie C., et al. "Active nonlocal metasurfaces." Nanophotonics 10.1 (2020): 655-665.
14. Zhou, You, et al. "Multiresonant nonlocal metasurfaces." Nano Letters 23.14 (2023): 6768-6775.
15. Yao, Jin, et al. "Nonlocal metasurface for dark-field edge emission." Science Advances 10.16 (2024): eadn2752.
16. Yao, Jin, et al. "Nonlocal Huygens' meta-lens for high-quality-factor spin-multiplexing imaging." Light: Science & Applications 14.1 (2025): 65.
17. Yao, Jin, et al. "Nonlocal meta-lens with Huygens' bound states in the continuum." Nature Communications 15.1 (2024): 6543.
18. Liang, Yao, Din Ping Tsai, and Yuri Kivshar. "From local to nonlocal high-Q plasmonic metasurfaces." Physical Review Letters 133.5 (2024): 053801.
19. Li, Guixin, Shuang Zhang, and Thomas Zentgraf. "Nonlinear photonic metasurfaces." Nature Reviews Materials 2.5 (2017): 1-14.
20. Krasnok, Alexander, Mykhailo Tymchenko, and Andrea Alù. "Nonlinear metasurfaces: a paradigm shift in nonlinear optics." Materials Today 21.1 (2018): 8-21.
21. Vabishchevich, Polina, and Yuri Kivshar. "Nonlinear photonics with metasurfaces." Photonics Research 11.2 (2023): B50-B64.
22. Sharma, Mukesh, et al. "Electrically and all-optically switchable nonlocal nonlinear metasurfaces." Science Advances 9.33 (2023): eadh2353.
23. Liu, Zhuojun, et al. "High-Q quasibound states in the continuum for nonlinear metasurfaces." Physical Review Letters 123.25 (2019): 253901.
24. Koshelev, Kirill, et al. "Nonlinear metasurfaces governed by bound states in the continuum." Acs Photonics 6.7 (2019): 1639-1644.
25. Sain, Basudeb, Cedrik Meier, and Thomas Zentgraf. "Nonlinear optics in all-dielectric nanoantennas and metasurfaces: a review." Advanced Photonics 1.2 (2019): 024002-024002.
26. Yang, Yuanmu, et al. "Nonlinear Fano-resonant dielectric metasurfaces." Nano letters 15.11 (2015): 7388-7393.
27. Reineke Matsudo, Bernhard, et al. "Efficient frequency conversion with geometric phase control in optical metasurfaces." Advanced Science 9.12 (2022): 2104508.

28. Tymchenko, Mykhailo, et al. "Gradient nonlinear pancharatnam-berry metasurfaces." Physical Review Letters 115.20 (2015): 207403.
29. Gennaro, Sylvain D., et al. "Nonlinear Pancharatnam–Berry phase metasurfaces beyond the dipole approximation." ACS Photonics 6.9 (2019): 2335-2341.
30. Mao, Ningbin, et al. "Nonlinear wavefront engineering with metasurface decorated quartz crystal." Nanophotonics 11.4 (2022): 797-803.
31. Liu, Bingyi, et al. "Nonlinear wavefront control by geometric-phase dielectric metasurfaces: influence of mode field and rotational symmetry." Advanced Optical Materials 8.9 (2020): 1902050.
32. Lin, Zemeng, et al. "Four-wave mixing holographic multiplexing based on nonlinear metasurfaces." Advanced Optical Materials 7.21 (2019): 1900782.
33. Jin, Boyuan, Dhananjay Mishra, and Christos Argyropoulos. "Efficient single-photon pair generation by spontaneous parametric down-conversion in nonlinear plasmonic metasurfaces." Nanoscale 13.47 (2021): 19903-19914.
34. Ma, Jinyong, et al. "Polarization engineering of entangled photons from a lithium niobate nonlinear metasurface." Nano Letters 23.17 (2023): 8091-8098.
35. Son, Changjin, et al. "Photon pairs bi-directionally emitted from a resonant metasurface." Nanoscale 15.6 (2023): 2567-2572.
36. Li, Lin, et al. "Metalens-array–based high-dimensional and multiphoton quantum source." Science 368.6498 (2020): 1487-1490.
37. Santiago-Cruz, Tomás, et al. "Resonant metasurfaces for generating complex quantum states." Science 377.6609 (2022): 991-995.
38. Zhang, Jihua, et al. "Spatially entangled photon pairs from lithium niobate nonlocal metasurfaces." Science advances 8.30 (2022): eabq4240.
39. Keren-Zur, Shay, et al. "Shaping light with nonlinear metasurfaces." Advances in Optics and Photonics 10.1 (2018): 309-353.
40. Gao, Jiannan, et al. "Topology-imprinting in nonlinear metasurfaces." Science Advances 11.24 (2025): eadv5190.
41. Chen, Shumei, et al. "High-order nonlinear spin–orbit interaction on plasmonic metasurfaces." Nano Letters 20.12 (2020): 8549-8555.
42. Coudrat, Laure, et al. "Unravelling the nonlinear generation of designer vortices with dielectric metasurfaces." Light: Science & Applications 14.1 (2025): 51.
43. Li, Zhi, et al. "Tripling the capacity of optical vortices by nonlinear metasurface." Laser & Photonics Reviews 12.11 (2018): 1800164.
44. Jeon, Dongmin, and Junsuk Rho. "Quasi-trapped guided mode in a metasurface waveguide for independent control of multiple nonlocal modes." Acs Photonics 11.2 (2024): 703-713.
45. Prokhorov, Alexei V., et al. "Design and tuning of substrate-fabricated dielectric metasurfaces supporting quasi-trapped modes in the infrared range." ACS Photonics 10.4 (2023): 1110-1118.
46. Evlyukhin, Andrey B., et al. "Polarization switching between electric and magnetic quasi-trapped modes in bianisotropic all-dielectric metasurfaces." Laser & Photonics Reviews 15.12 (2021): 2100206.
47. Sayanskiy, Andrey, et al. "Controlling high-Q trapped modes in polarization-insensitive all-dielectric metasurfaces." Physical Review B 99.8 (2019): 085306.'
48. Alaee, Rasoul, Carsten Rockstuhl, and Ivan Fernandez-Corbaton. "An electromagnetic multipole expansion beyond the long-wavelength approximation." Optics Communications 407 (2018): 17-21.
49. Alaee, Rasoul, Carsten Rockstuhl, and Ivan Fernandez-Corbaton. "Exact multipolar decompositions with applications in nanophotonics." Advanced Optical Materials 7.1 (2019): 1800783.
50. Evlyukhin, Andrey B., and Boris N. Chichkov. "Multipole decompositions for directional light scattering." Physical Review B 100.12 (2019): 125415.

51. Rodriguez, Alejandro, et al. "χ (2) and χ (3) harmonic generation at a critical power in inhomogeneous doubly resonant cavities." Optics Express 15.12 (2007): 7303-7318.
52. Molesky, Sean, et al. "Inverse design in nanophotonics." Nature Photonics 12.11 (2018): 659-670.
53. Noor, Ahsan, et al. "Mode-matching enhancement of second-harmonic generation with plasmonic nanopatch antennas." ACS photonics 7.12 (2020): 3333-3340.
54. Gigli, Carlo, et al. "Quasinormal-mode non-hermitian modeling and design in nonlinear nano-optics." ACS photonics 7.5 (2020): 1197-1205.
55. Gao, Yisheng, et al. "Nonlinear holographic all-dielectric metasurfaces." Nano letters 18.12 (2018): 8054-8061.
56. Ye, Weimin, et al. "Spin and wavelength multiplexed nonlinear metasurface holography." Nature Communications 7.1 (2016): 11930.
57. Frese, Daniel, et al. "Nonlinear bicolor holography using plasmonic metasurfaces." ACS photonics 8.4 (2021): 1013-1019.
58. Almeida, Euclides, Ora Bitton, and Yehiam Prior. "Nonlinear metamaterials for holography." Nature Communications 7.1 (2016): 12533.
59. Mao, Ningbin, et al. "Nonlinear vectorial holography with quad-atom metasurfaces." Proceedings of the National Academy of Sciences 119.22 (2022): e2204418119.
60. Jiang, Qiang, Guofan Jin, and Liangcai Cao. "When metasurface meets hologram: principle and advances." Advances in Optics and Photonics 11.3 (2019): 518-576.
61. Ouyang, Xu, et al. "Ultra-narrowband geometric-phase resonant metasurfaces." Proceedings of the National Academy of Sciences 122.15 (2025): e2420830122.
62. Yu, Nanfang, et al. "Light propagation with phase discontinuities: generalized laws of reflection and refraction." Science 334.6054 (2011): 333-337.
63. Kim, Yeseul, et al. "Spin-Dependent Phenomena of Meta-optics." ACS Photonics 12.1 (2024): 16-33.
64. Jisha, Chandroth Pannian, Stefan Nolte, and Alessandro Alberucci. "Geometric phase in optics: from wavefront manipulation to waveguiding." Laser & Photonics Reviews 15.10 (2021): 2100003.
65. Karnieli, Aviv, Yongyao Li, and Ady Arie. "The geometric phase in nonlinear frequency conversion." Frontiers of Physics 17.1 (2022): 12301.